\documentclass[aps,prl,preprint,groupedaddress,superscriptaddress,floatfix,12pt]{revtex4-1}

\usepackage{graphicx}

\begin{document}

\title{Multiple Silicon Atom Artificial Molecules}

\author{John A. Wood}

\affiliation{Department of Physics, University of Alberta, Edmonton, Alberta, T6G 2J1, Canada.}

\author{Mohammad Rashidi}
\affiliation{Department of Physics, University of Alberta, Edmonton, Alberta, T6G 2J1, Canada.}
\affiliation{National Institute for Nanotechnology, National Research Council of Canada, Edmonton, Alberta, T6G 2M9, Canada.}

\author{Mohammad Koleini}
\affiliation{Department of Physics, University of Alberta, Edmonton, Alberta, T6G 2J1, Canada.}
\affiliation{National Institute for Nanotechnology, National Research Council of Canada, Edmonton, Alberta, T6G 2M9, Canada.}

\author{Jason L. Pitters}
\affiliation{National Institute for Nanotechnology, National Research Council of Canada, Edmonton, Alberta, T6G 2M9, Canada.}

\author{Robert A. Wolkow}
\email{rwolkow@ualberta.ca}
\affiliation{Department of Physics, University of Alberta, Edmonton, Alberta, T6G 2J1, Canada.}
\affiliation{National Institute for Nanotechnology, National Research Council of Canada, Edmonton, Alberta, T6G 2M9, Canada.}

\date{\today}

\begin{abstract}
We present linear ensembles of dangling bond chains on a hydrogen terminated Si(100) surface, patterned in the closest spaced arrangement allowed by the surface lattice. Local density of states maps over a range of voltages extending spatially over the close-coupled entities reveal a rich energetic and spatial variation of electronic states. These artificial molecules exhibit collective electronic states resulting from covalent interaction of the constituent atoms. A pronounced electrostatic perturbation of dangling bond chain structure is induced by close placement of a negatively dangling bond.  The electronic changes so induced are entirely removed, paradoxically, by addition of a second dangling bond.
\end{abstract}


\maketitle
\section{Introduction}
Three-coordinate silicon atoms have associated dangling bond (DB) states.  Discovery of the interactions, distinct charge states, geometric and electrostatic control over occupation and polarization among other properties led to the recognition that such states can be viewed as atomic silicon quantum dots, ASiQDs \cite{haider_prl2009}.  Like other quantum dots the ASiQDs can be deployed as the building blocks with such structures as quantum cellular automata, spin and charge qubits, and various other passive and active entities~ \cite{Livadaru2010a,Shaterzadeh-Yazdi2014,Wolkow2014}. 

Through combined experimental and theoretical modeling investigations~\cite{Livadaru2011, Pitters2011}, including studies of single atom, single electron dynamics~\cite{Taucer2014,labidi_njp2015,Rashidi2015} many properties of the ASiQDs have been revealed.

Closest spaced DB structures, with no intervening H-terminated Si sites between DBs, were first studied by Hitosugi et al.~\cite{Hitosugi1999}.  It was there concluded that DBs positioned along one side of a dimer row on the Si(100) surface were subject to a Jahn-Teller distortion.  Consequently, electronic states measured by scanning tunneling microscopy (STM) showed maxima not simply related to the geometric position of constituent atoms. They however did not explore a range of energies and as a result missed the diverse electronic structural character evident when a wide range of energies are explored, as shown here.  Engelund et al. have studied DB pairs with this separation~\cite{Engelund2015}.  Schofield et al. showed DB structures in the weak coupling regime with hydrogen terminated Si atoms intervening between DBs~\cite{schofield_nc2013}.

The strong coupling regime studied here is only evident when DBs are as close spaced as the lattice allows.  That spacing is 3.84~\AA.  We have created and studied such close-spaced structures on the H-Si(100) and H-Si(111) hydrogen atom terminated silicon surfaces.  The former will be described in this work.

Going beyond the initial study by~\cite{Hitosugi1999}, we describe 3, 4, 5, 6, and 7 atom close-space linear ensembles of ASiQDs on the H-Si(100) surface.  Occupied and unoccupied state STM images were recorded for each.   Most revealingly, d\textit{I}/d\textit{V} maps over a range of voltages extending spatially along the axis of the artificial molecule reveal the rich energetic and spatial variation of electronic states.  It is clear that, unlike groups of DBs with intervening H-terminated silicon sites, the close couple entities examined here show true collective electronic states resulting from covalent interaction of the constituent atoms.  While ab initio modeling is not included in this work it is clear that the artificial molecules, or what we may alternatively describe as extended quantum dots, show electronic structure that is not linearly derived from the sum of the atomic components.  It is clear too that the local density of states observed are highly convolved with tip induced band bending effects and as a result a simple association of observed images with unperturbed electronic structure cannot be made - assistance from theory is required.

A pronounced electrostatic perturbation of the 7 DB structure is induced by close placement of a negatively charged DB.  The electronic changes so induced are entirely removed, paradoxically, by addition of a second DB.  The second DB was formed on the same underlying silicon dimer.  Two such closed placed dimers, approximately 2.4~\AA removed, are known to form $\pi$ and $\pi^*$ bonds which are respectively aligned energetically with the valence and conduction band edges, preventing charge localization.

Coupling of atomic dots is not so large, of order 100~meV, as to leave new collective states outside of the band gap.  The near proximity of a H-free dimer as described above provides a way to substantially enhance the electrical connection of the multi atom quantum dots described here to the bulk silicon.  Structures of the kind defined here will have wide application in the atom scale classical and quantum electronic circuitry soon to emerge~\cite{Livadaru2010a}.

\section{Experiments}
DBs can be created controllably with the probe of a scanned probe microscope.  A voltage and or current larger than is used for imaging is applied briefly via the probe to the target hydrogen atom to be removed.  A voltage is chosen such that current rises at that voltage when an hydrogen atom is removed.  Upon checking for that change in current and finding a hydrogen removal the electrical conditions applied to break the targeted Si-H bond are ceased.  Re-imaging reveals the newly created DB.  Patterns can be created by placing the tip over multiple desired positions to create multiple DBs. We developed an algorithm to automate this process. In this way, two to seven long DB chains were patterned along and on the same side of a dimer row in the closest spaced arrangement allowed by the lattice (0.384~nm). 

A lock-in amplifier collected the d\textit{I}/d\textit{V} signal at each point. Thus we mapped out the local density of states, LDOS, of the DB chain as a function of voltage (usually from -0.4~V to -1.85~V). We note that our experimental technique is not afflicted by z-drift (tip sample separation) as the tip height is reset each time a line or map is completed and drift during the time taken to move along a line or scan over a map, a maximum of 3 minutes, is negligible.

Experiments were conducted in an Omicron Low Temperature STM at 4.5~K under ultra high vacuum (UHV). An lock-in amplifier was used to measure d\textit{I}/d\textit{V} signal (modulation frequency of 760--820~Hz and amplitude of 30~mV). 

Arsenic doped (0.002--0.003 m$\Omega$/cm) Si(100) samples were direct current heated to 1050$^\circ$C for a short time for oxide desorption, and hydrogen terminated at 330$^\circ$C for $\approx$20s under hydrogen exposure, forming the H-Si(100) 2$\times$1 surface. It is known that flashing to 1050$^\circ$C does not significantly remove dopants in the near-surface regime~\cite{Pitters2012} and so a uniform dopant profile persists all the way to the surface.
 
Polycrystalline electrochemically etched tungsten tips were heated to ~800$^\circ$C for about two minutes under UHV condition for cleaning and oxide desorption. Their quality was checked by field ion microscope (FIM), and nitrogen etched to obtain single atom tips~\cite{Rezeq2006}. Small tip modifications were made during STM measurements by slightly contacting the tip with a patch of bare Si while applying a voltage of -2.0 to -3.0~V.

\section{Results}

\begin{figure}
\includegraphics{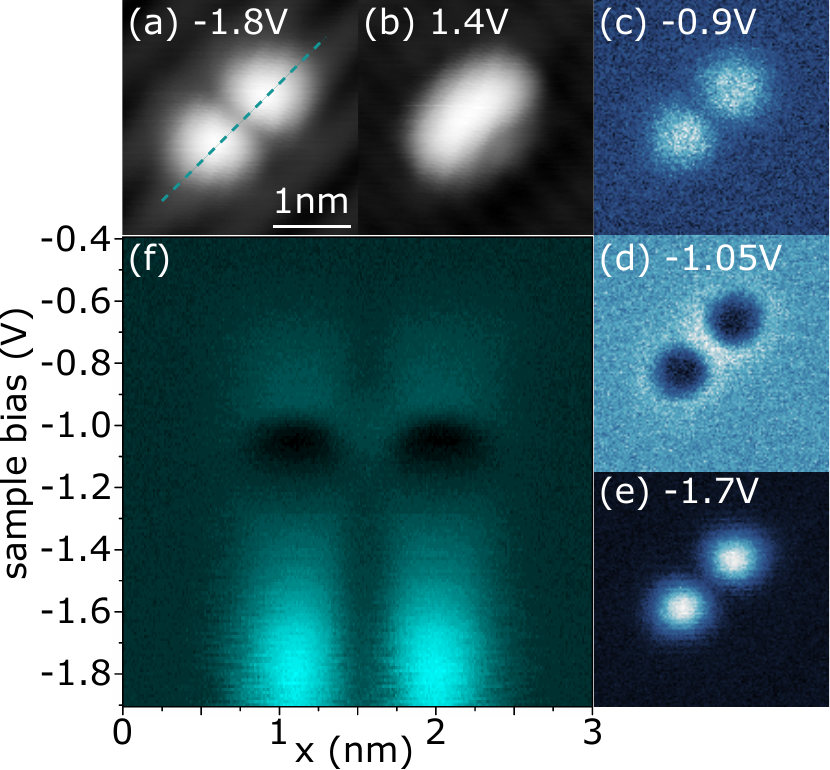}
\caption{\label{Fig1} 3DB Chain. (a,b) STM constant current images at -1.8V and 1.4V respectively. The current set point was 50pA. (c--e) d\textit{I}/d\textit{V} maps taken at a tip height set at -1.8~V, 20~pA with a 60~pm tip retraction over an H:Si dimer. The sample bias during collection of each d\textit{I}/d\textit{V} map is labeled in the upper left corner of the map. The scale bar for the STM images and the d\textit{I}/d\textit{V} maps is shown in (a). (f) d\textit{I}/d\textit{V} line scans at varying energy along the 3nm dotted line overlaid on the STM image (a). }
\end{figure}

Figure~\ref{Fig1} shows various characterizations of a linear 3 DB chain formed along one side of a dimer row on the Si(100) H terminated surface.

The large d\textit{I}/d\textit{V} vs position vs sample bias graphic, Fig.~\ref{Fig1}(f) conveys much information about the energy and spatial distribution of local state density.  In other words, it shows where electrons are localized, and where electrons are diminished, at each energy probed.

At most energies, strikingly, we see little state density at the central atom position.  This is a substantial departure from what might be expected by simply summing the contributions of three non-interacting ASiQDs and a clear indication of newly emergent molecule like spectroscopy of the ensemble resulting from quantum mechanical overlap of atomic like orbitals of the constituents.

Further to the character of the 3 DB chain,  Fig.~\ref{Fig1}(a) and (b) shows STM constant current images collected at -1.8~V and 1.4~V respectively. The set point current was 50~pA. In filled states, the 3 DB chain appears as two bright spots. In empty states, the 3 DB Chain appears as one bright feature spread over the length of the chain. Figure~\ref{Fig1}(c--e) show constant height d\textit{I}/d\textit{V} maps taken of the 3 DB chain at three different bias voltages: -0.9~V, -1.05~V, and -1.7~V. Note that the color scale for the three d\textit{I}/d\textit{V} maps is different for each map. Figure~\ref{Fig1}(f) shows a plot of d\textit{I}/d\textit{V} line-scans along the axis of the 3 DB chain, shown in a dotted line in (a), from -0.4~V to -1.9~V with 10~mV resolution. For (c--f), the tip height was set at -1.8~V and 20~pA over an H-Si dimer near the chain with 60~pm tip retraction after the feedback loop was turned off. We observe a change of the patterns in d\textit{I}/d\textit{V} maps and linescans as the voltage is increased. We identify three distinct regions in the d\textit{I}/d\textit{V} maps of the 3 DB chain: From -0.6~V to -1.0~V they take on the appearance of two bright spots. From -1.0~V to -1.2~V they take on the appearance of two dark spots in the same spatial location. We observe negative differential resistance within the spots, with two bright rings that surround the spots. Finally, From -1.2~V to -1.9~V we observe two bright spots once again.

\begin{figure*}
\includegraphics[width=11cm]{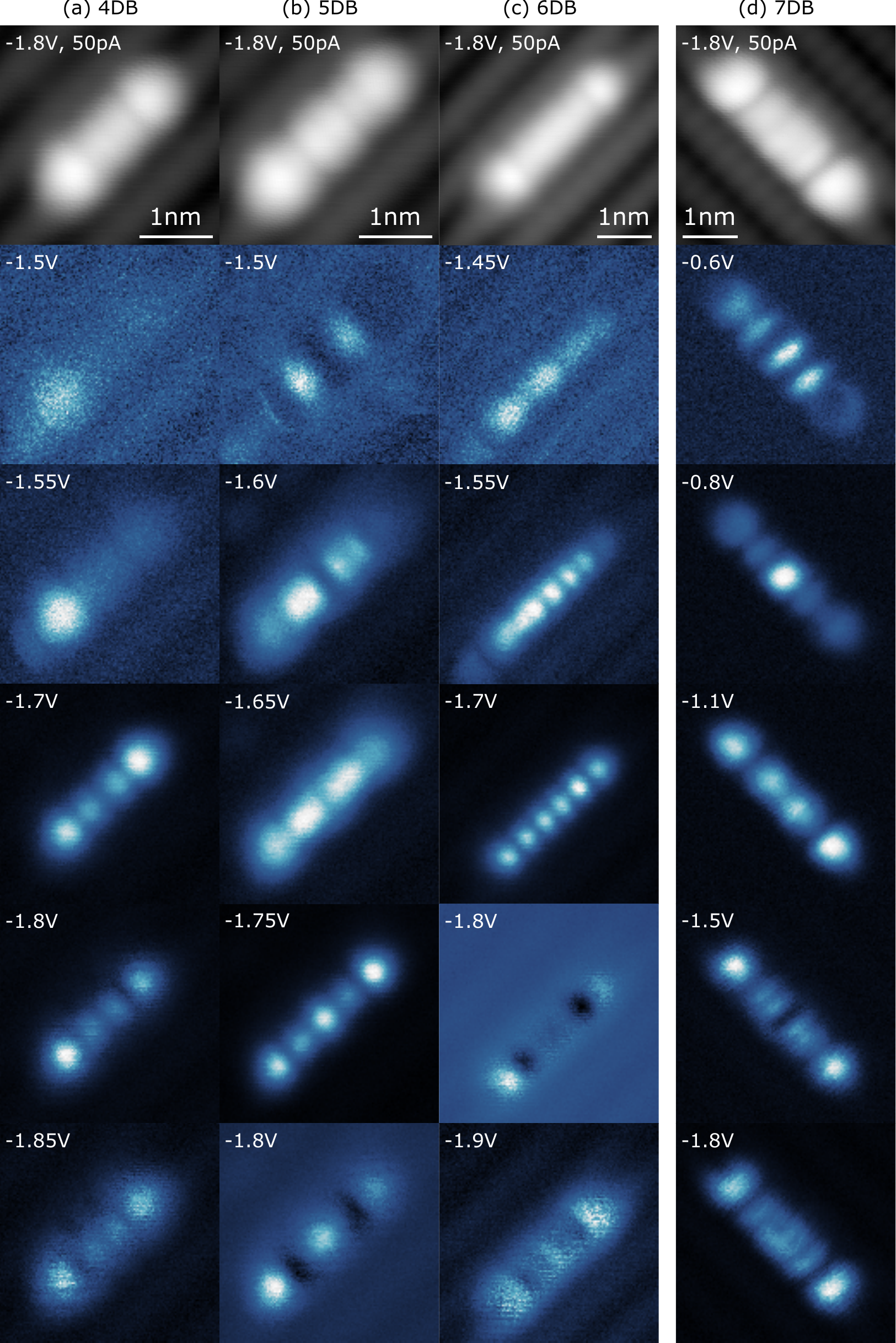}
\caption{\label{Fig2} DB Chains of greater length. (a) 4 DB Chain, (b) 5 DB Chain, (c) 6 DB Chain, (d) 7 DB Chain. The 5 DB (b) and 6 DB (c) Chains were formed by adding DBs to the upper right end of the 4 DB chain (a). The first image of each column is an STM constant current image at -1.8~V, 50~pA. All other images are d\textit{I}/d\textit{V} maps taken at a tip height of -1.8~V, 20~pA with a 60~pm tip retraction over an H-Si region. The sample bias during collection of each d\textit{I}/d\textit{V} map is labeled in the upper left corner of the map. The scale bar for both the STM image and the d\textit{I}/d\textit{V} maps for each column is shown at the bottom of each column’s STM image. }
\end{figure*}

Figure~\ref{Fig2} shows results for a variety of other chain lengths.  In all cases there are no Si-H entities in between the DBs.  Rather, the DBs are as close together as the silicon lattice allows, 3.84~\AA.

In Fig.~\ref{Fig2} it is seen that diverse electronic properties are observed as a function of number of constituent atoms, as well as of functions of energy and position. 

In the d\textit{I}/d\textit{V} maps of DB chains of length 4,5,6, and 7 in Fig.~\ref{Fig2} each column corresponds to one particular DB Chain length. The first image of each column is an STM constant current image of that chain, imaged at -1.8V with a set point current of 50~pA. The remaining images are d\textit{I}/d\textit{V} maps of that chain taken at a tip height of -1.8~V, 20~pA with a tip retraction of 60~pm over a H-Si dimer. We note that for all chains the patterns in d\textit{I}/d\textit{V} maps change as a function of sample bias. The bright spots often do not correspond to the locations of the DBs. Thus the patterns must be emergent from an interaction between the DBs that make up the chain. Most STM images and d\textit{I}/d\textit{V} maps are symmetric about the center of the DB chain. The 5 and 6 DB chains in (b,c) were made by extending the 4 DB chain in (a). Knowing this, we note that similar features appear at similar voltages for the 4,5, and 6 DB chains shown: At a sample bias between -1.7~V to -1.75~V, the d\textit{I}/d\textit{V} maps show the same number of bright spots as the number of DBs in the chain. For the 5 and 6DB chain at a sample bias of -1.8~V, the d\textit{I}/d\textit{V} maps appear to show spots with negative differential resistance.

\begin{figure*}
\includegraphics{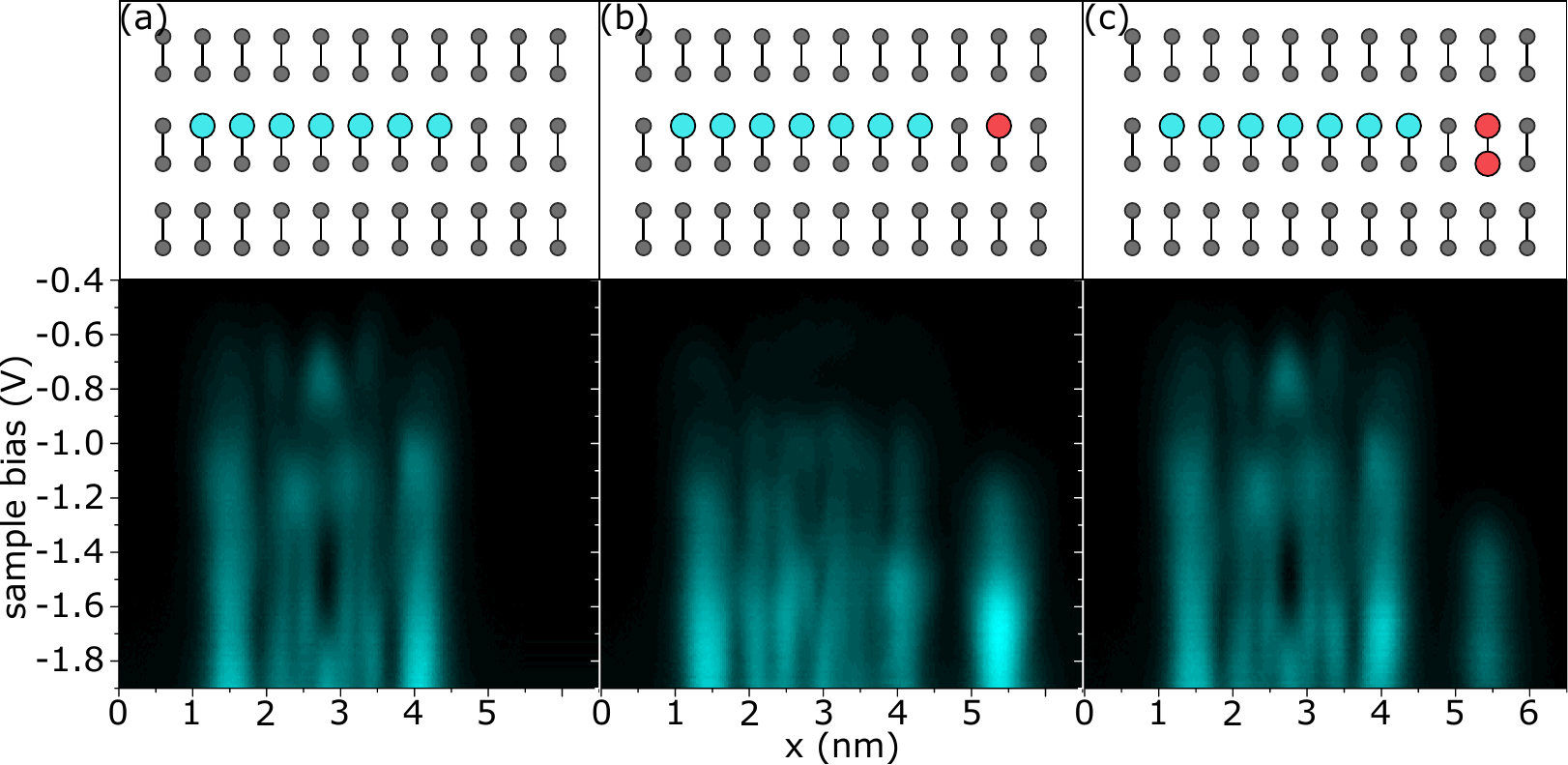}
\caption{\label{Fig3} Controlled perturbation of a 7 DB chain (a) with a single DB (b) and with a bare dimer (c). The top images in (a-c) are diagrams of the surface DB structures studied. Grey circles indicate H-Si surface atoms. Larger circles indicate Si DBs patterned by the STM tip. The light blue colored circles are DBs that make up the 7 DB chain. The red colored circles are DBs that make up the perturbing surface feature. The lines connecting pairs of surface features represent the bonds between the underlying Si atoms - the dimers of the H-Si(100) 2$\times$1 surface. The bottom images are d\textit{I}/d\textit{V} linescans of the 7 DB chain and the perturbing feature along the central axis of the 7 DB chain. The height setpoint of the d\textit{I}/d\textit{V} linescans was -1.8~V, 20~pA with a tip retraction of 60~pm over a H-Si dimer. The x scale of each d\textit{I}/d\textit{V} linescan is lined up with the x scale of each diagram. The perturbing single DB in (b) is located on the same side of the dimer row as the 7 DB chain, with one intervening H-Si dimer between the DB and the 7 DB chain. The perturbing bare dimer in (c) was made by creating a DB on the same dimer as the original perturbing DB in (b). In all d\textit{I}/d\textit{V} linescans, the state density associated with the 7 DB chain is found at an x position of 1nm to 4.5nm.  In the d\textit{I}/d\textit{V} linescans of (b) and (c), the the state density of the single DB and the bare dimer respectively are found at an x position of ~5.5nm. }
\end{figure*}

Figure~\ref{Fig3} shows the electrostatic perturbation effect of a localized charge on a 7 DB chain. The  d\textit{I}/d\textit{V} line-scans of the unperturbed chain (Fig.~\ref{Fig3}(a) exhibit symmetry about the center of the DB chain. A significant d\textit{I}/d\textit{V} signal begins at a sample bias of about -0.6~V.

In Fig.~\ref{Fig3}(b), we see the same 7 DB chain perturbed by a single DB on the same side of the dimer row with one intervening H-Si dimer.  The d\textit{I}/d\textit{V} linescans show higher state density on the end of the chain farthest from the single DB. They show patterns very different from that of the unperturbed 7DB chain. Although a faint d\textit{I}/d\textit{V} signal is seen from -0.6~V to -0.9~V, it is substantially less than the magnitude of the signal seen at the same voltages on the unperturbed 7 DB chain.

Another DB was created adjacent to the initial perturbing DB on the same Si Dimer, creating a bare dimer. The d\textit{I}/d\textit{V} linescans of Fig.~\ref{Fig3}(c) show that this has the effect of returning the 7 DB chain to its unperturbed character (Fig.~\ref{Fig3}(a)).

\section{Discussion}
When hydrogen atoms intervene between DBs in extended chains, as is the case in all but one previous published examples\cite{Hitosugi1999}, the strong mixing and binding of atomic states does not occur and the qualities described here are not observed.  When intervening hydrogen atoms exist between DBs the images of such widely spaced ensembles reveal only a simple sum of the qualities of the constituent parts -not as seen here- newly emergent properties that are non-linearly related to the properties of the constituents.

Close spaced DBs form substantial covalent (electron sharing) bonds.  The newly emergent electronic structure is observed in multiple modes of STM imaging.  Spatially point specific d\textit{I}/d\textit{V} spectra show pronounced changes indicating new electronic structure.  2D maps of d\textit{I}/d\textit{V}, taken at a specific V show various spatial density of states variations at that energy across the area of the collective.  d\textit{I}/d\textit{V} spectra over a range of voltages, taken along a line from one end of a molecule to the other, reveal a rich spectral map of the molecule’s density of states.  Such views make it compelling to describe the ensembles as new artificial molecules.

\begin{figure}
\includegraphics{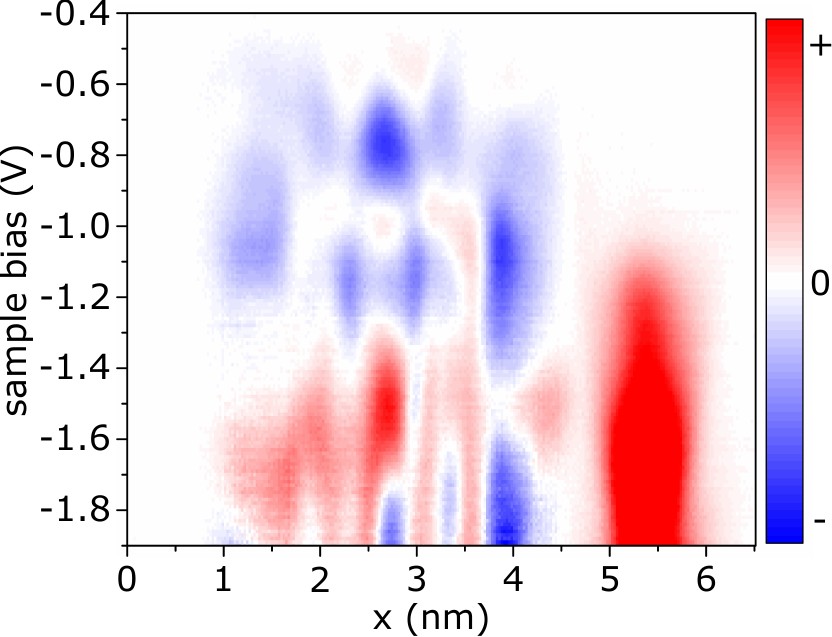}
\caption{\label{Fig4} A difference map, derived from Fig.~\ref{Fig3}(a,b), showing the altered electronic structure resulting from an electrostatic perturbation.  Red corresponds to increased state density and blue to a reduction. }
\end{figure}

The observations in Fig.~\ref{Fig3} show that by placing a DB, or a localized charge, next to the 7 DB chain, we electrostatically influence the chain and cause its electron state density to shift away from the perturbing localized charge. Fig.~\ref{Fig4} shows a difference map, derived from Fig.~\ref{Fig3}(a,b), showing the change in electronic structure resulting from an electrostatic perturbation.  Red corresponds to increased state density and blue to a reduction.  The localized charge also increases the energy of all electrons on the chain, causing a great reduction of state density at lower voltages, -0.6~V to -0.9~V, compared with the unperturbed 7 DB chain.  Similar Stark shifts were seen for very different molecules near a charged DB.  In that case, molecules from a bottle, not artificial molecules were studied~\cite{Piva2005}. 

A single DB naturally attains a negative charge on a highly doped substrate as is used here.  It is that localized charge that is electrostatically altering the properties of the adjacent molecule.

Somewhat surprisingly, creating another DB immediately adjacent to the first perturbing DB, specifically on the same underlying Si dimer, eliminates the negative localized charge.  This effect occurs because the strongly interacting nature of 2 DBs on one dimer causes a large energetic splitting.  So much so that the new symmetric and antisymmetric states created – often referred to as $\pi$ and $\pi^*$ states, are resonant respectively with the bulk silicon valence and conduction bend edges.  As a result, the electrons do not localize but rather are disbursed in those bands.  

Upon replacing the single perturbing DB with 2 DBs of a bare Si dimer it is evident that the d\textit{I}/d\textit{V} linescans of the 7 DB chain have returned to their unperturbed state.

In addition to showing the reversibility of the electronic perturbation effect on the multi silicon atom artificial molecule, this effect demonstrates the utility of a clean dimer as a way to connect gap states such and ensembles thereof to bulk states.


%

\end{document}